\documentclass[12pt]{article}
\usepackage{latexsym}
\usepackage{amssymb,amsfonts,amsmath}
\usepackage{graphicx} 
\usepackage{indentfirst}
\usepackage{bbm}
\usepackage{amssymb}
\usepackage{verbatim}
\usepackage{amsmath, amsthm,amssymb}
\usepackage{mathrsfs}
\usepackage{hyperref}
\usepackage{amsfonts}
\usepackage{dsfont}

\newcommand {\cD}{{\cal D}}
\newcommand {\cE}{{\cal E}}

\newcommand {\cH}{{\cal H}}

\newcommand {\cM}{{\cal M}}
\newcommand {\cN}{{\cal N}}

\newcommand {\cV}{{\cal V}}

\newcommand {\cX}{{\cal X}}
\newcommand {\cY}{{\cal Y}}
\newcommand {\cZ}{{\cal Z}}

\def\a{\alpha}
\def\b{\beta}

\def\d{\delta}

\def\g{\gamma}

\def\j{\psi}
\def\k{\kappa}

\def\m{\mu}

\def\q{\theta}
\def\r{\rho}

\def\t{\tau}

\def\z{\zeta}
\def\D{\Delta}
\def\F{\Phi}
\def\J{\Psi}
\def\L{\Lambda}
\def\O{\Omega}

\def\S{\Sigma}

\def\ri{{\rm i}}
\def\re{{\rm e}}

\newcommand{\ad}{{\dot{\alpha}}}                           
\newcommand{\bd}{{\dot{\beta}}}                            
\newcommand{\ve}{\varepsilon}                            
\newcommand{\cDB}{{\bar\cD}}                            

\newcommand{\pa}{\partial}                           
\newcommand{\hf}{\frac12}

\newcommand{\be}{\begin{equation}}
\newcommand{\ee}{\end{equation}}
\newcommand{\bea}{\begin{eqnarray}}
\newcommand{\eea}{\end{eqnarray}}
\newcommand{\non}{\nonumber}
\newcommand{\ba}{\begin{array}}
\newcommand{\ea}{\end{array}}

\def\double #1{#1{\hbox{\kern-2pt $#1$}}}

\newcommand{\gd}{{\dot\g}}

\newcommand{\sSU}{\mathsf{SU}}

\newcommand{\bsubeq}{\begin{subequations}}
\newcommand{\esubeq}{\end{subequations}}

\newcommand{\rd}{\mathrm d}
\numberwithin{equation}{section}

\begin{document}

\begin{center}
{\Large \bf 
Goldstino superfields in supergravity}
\end{center}

\begin{center}
{\bf Sergei M. Kuzenko} \\
\vspace{5mm}

\footnotesize{
{\it Department of Physics M013, The University of Western Australia\\
35 Stirling Highway, Crawley W.A. 6009, Australia}}  
~\\
\vspace{2mm}
\end{center}

\begin{abstract}
\baselineskip=14pt
We review two off-shell models  for spontaneously broken 
$\cN=1$ and $\cN=2$ supergravity  proposed in arXiv:1702.02423
and arXiv:1707.07390. The $\cN=1$ model makes use of a real scalar 
superfield subject to three nilpotency conditions. The $\cN=2$ theory is
formulated in terms of a reduced chiral superfield obeying a cubic nilpotency condition.
New results on nilpotent $\cN =1$ supergravity are also
included.
\end{abstract}

\renewcommand{\thefootnote}{\arabic{footnote}}


\section{Introduction}

According to the general relation between linear and nonlinear realisations of 
$\cN=1$ supersymmetry established by Ivanov and Kapustnikov  
\cite{IK}, 
the Volkov-Akulov Goldstone fermion (Goldstino) \cite{VA} 
may equivalently be described in terms of  a constrained superfield. 
Such a Goldstino superfield is called 
 irreducible \cite{BHKMS} since  the Goldstino is 
 its only independent component.
 There also exist reducible Goldstino superfields that contain auxiliary field(s) 
 in addition to the Goldstino.
 The very first example of an irreducible Goldstino superfield in four dimensions
 was the nilpotent chiral scalar $\cX$ introduced in  \cite{IK,Rocek}.
Ro\v{c}ek  \cite{Rocek} defined
 $\cX$, $\bar D_\ad \cX=0$, to obey the nilpotency condition $\cX^2=0$ 
 and nonlinear constraint ${f}\cX = -\frac 14  \cX \bar D^2 \bar \cX$,
 where $f$ is a real parameter 
  characterising the scale of supersymmetry breaking, 
 and $D_A =(\pa_a , D_\a, \bar D^\ad)$ are the covariant derivatives of 
$\cN=1$ Minkowski superspace.
 The first reducible Goldstino superfield was proposed by Casalbuoni {\it et al.}
 \cite{Casalbuoni} and rediscovered, in a different framework, 
 by Komargodski and Seiberg \cite{KS}. 
 It is a chiral scalar $X$ subject to the only constraint $X^2=0$.
 As argued in \cite{BHKMS},
 every reducible Goldstino superfield may always be represented
as an irreducible one plus a `matter' superfield, which contains
all the component fields except for the Goldstino. For instance, 
it was shown in \cite{BHKMS} that 
\bea
X = \cX +\cY~,\qquad 
{f} \cX := - \frac{1}{4} \bar D^2 (\bar \S \S)\, , \qquad 
\S :=  - 4 f \frac{\bar X}{\bar D^2 \bar X}~,
\label{1.1}
\eea
where the auxiliary field $F$ of $X$ is the only independent component of the 
chiral scalar $\cY$. Originally, 
the irreducible Goldstino  superfield  $\S$
 was introduced in \cite{KTyler} to be a
modified complex linear superfield, $-\frac 14 {\bar D}^2\Sigma={f}$, 
which is nilpotent, $\S^2=0$, and obeys the holomorphic nonlinear constraint
${f}D_\alpha\Sigma = -\frac 14 \Sigma{\bar D}^2D_\alpha\Sigma$.
These properties follow from \eqref{1.1}.

Due to the universality of the Volkov-Akulov action \cite{IK,VA},
all irreducible Goldstino superfield models are equivalent.
There exists a computer program created by Tyler \cite{KT12} 
to construct the most general (twelve-parameter) field redefinition 
that relates any Goldstino model to the Volkov-Akulov action. 
Explicit relations  that express every irreducible Goldstino superfield 
in terms of a given one have been worked out in \cite{BHKMS,KTyler,SW,CDF}.
All irreducible Goldstino superfields share one remarkable feature
discovered in \cite{BK17}. Each of them
may be realised as a composite of $X$ and its conjugate $\bar X$ 
that is invariant under arbitrary  local rescalings 
$X \to \re^{\t} X$, with the parameter $\t$ being chiral;
an example is provided by $\S$ given by eq. \eqref{1.1}.
This result implies, in fact, that any coupling of $X$ to a supergravity-matter system
is dynamically equivalent to the same system coupled to a nilpotent chiral scalar $\cX$, 
$\cX^2=0$, subject to a suitable deformation of the constraint  
$f \cX = -\frac 14  \cX \bar D^2 \bar \cX$, see also \cite{BHKMS}.

When a Goldstino superfield is coupled to off-shell supergravity, the local supersymmetry becomes spontaneously broken, in accordance with the super-Higgs 
effect \cite{VS,DZ}.
This is accompanied by the appearance of a positive contribution 
to the cosmological constant, which is proportional to $f^2$. 
The latter phenomenon was first  observed in 1977 
by Deser and Zumino  within on-shell supergravity \cite{DZ},
and a year later by Lindstr\"om and Ro\v{c}ek \cite{LR} 
who constructed the first off-shell model for spontaneously broken local $\cN=1$ supersymmetry in four dimensions. 
They coupled the nilpotent chiral scalar $\cX$  of \cite{Rocek}
to old minimal supergravity, with a supersymmetric cosmological term included. 
Their work completed the earlier attempt made  in \cite{DZ}
to couple the Volkov-Akulov action \cite{VA} to supergravity.\footnote{Since Deser 
and Zumino \cite{DZ} made use of on-shell supergravity, 
it was next to impossible to construct a 
complete supergravity-Goldstino action in their setting.} 
The coupling of $X$ to old minimal supergravity was worked out in detail 
by two groups in 2015  \cite{BFKVP,HY}.
The  work by Bergshoeff {\it et al.}  \cite{BFKVP} put forward the concept of de Sitter supergravity, which has renewed interest in spontaneously broken supergravity.

This paper is a review of two  models for spontaneously broken $\cN=1$ and $\cN=2$
supergravity proposed in recent publications \cite{KMcAT-M,KT-M17}.


\section{Nilpotent $\cN=1$ real scalar multiplet}

The $\cN=1$ Goldstino superfield model proposed in \cite{KMcAT-M} 
is described in terms  
of a real scalar superfield $V$ subject to 
the three nilpotency constraints\footnote{In curved superspace, the
covariant derivatives $\cD_A = (\cD_a, \cD_\a, \bar \cD^\ad)$ 
have the form
$\cD_{A}= E_A{}^M\pa_M+\hf \O_A{}^{bc}M_{bc}$,
where $M_{bc}$ is the Lorentz generator.
Our description of the old minimal formulation for $\cN=1$ supergravity follows 
\cite{Ideas}, where 
the graded commutation relations for the covariant derivatives are given.} 
\begin{subequations} \label{2.1}
\bea
V^2&=&0~, \label{2.1a}\\
V \cD_A \cD_B V &=&0~,\label{2.1b} \\
V \cD_A \cD_B \cD_C V &=&0~. \label{2.1c}
\eea
\end{subequations}
It is also necessary to require that the real descendant 
$\cD W:=\cD^\a W_\a =\bar \cD_\ad \bar W^\ad$ be nowhere vanishing, with
\bea
W_\a := -\frac{1}{4} (\bar \cD^2 - 4R) \cD_\a V~.
\label{2.2}
\eea
Here  $R$  is one of the 
torsion tensors $R$, $G_a = {\bar G}_a$ and
$W_{\a \b \g} = W_{(\a \b\g)}$ 
which determine the curved superspace geometry 
(see  \cite{Ideas} for a review), 
with $R$ and $W_{\a\b\g}$ being covariantly chiral.
Because of the constraints imposed, 
 $V$ has only two independent component fields, 
the Goldstino $\j_\a \propto W_\a|_{\q=0}$ and the auxiliary scalar $D\propto \cD^\a W_\a|_{\q=0}$. 
As shown in \cite{KMcAT-M},
the constraints \eqref{2.1} imply the representation
\bea
V = - 4 \frac{W^2 \bar W^2}{(\cD W)^3}~, \qquad 
W^2 := W^\a W_\a~, 
\label{2.3}
\eea
which ensures the fact that the constraints \eqref{2.1} hold.
The dynamics of this supermultiplet is governed by the super-Weyl invariant action
\bea
S[V] =   \int \rd^4 x \rd^2 \q  \rd^2 \bar{\q} \, E\,\Big\{
\frac{1}{16} V \cD^\a (\bar \cD^2 -4R ) \cD_\a V- 2f \bar \F \F V\Big\}~,
\label{2.4}
\eea
where  
$\F$ is the chiral compensator, $\bar \cD_\ad \F =0$, 
for the old minimal formulation for $\cN=1$ supergravity, 
and $E^{-1} = \text{Ber} (E_A{}^M)$.

The constraints \eqref{2.1} are invariant under local re-scalings of $V$,
$V  \to  \re^\r V$,
with $\r$ an arbitrary real scalar superfield. Requiring the action \eqref{2.4} to be stationary 
under such rescalings leads to the nonlinear constraint 
\bea
f \bar \F \F V =\frac{1}{16} V \cD^\a (\bar \cD^2 -4R ) \cD_\a V~,
\label{2.5}
\eea
which expresses the auxiliary scalar $D$ in terms of the Goldstino.
The set of constraints \eqref{2.1} and \eqref{2.5} defines the irreducible Goldstino superfield $\cV$  introduced in \cite{BHKMS}.
The constraints \eqref{2.1a} and  \eqref{2.2} appeared originally 
in \cite{LR}, and later were discussed in  \cite{SW}.
In both papers \cite{LR,SW}, the Goldstino superfield $\cV$ was considered as a composite 
superfield, $f \cV = \bar \cX \cX$. However, if $\cV$ is viewed as a fundamental 
Goldstino superfield, then the constraints  \eqref{2.1b} and  \eqref{2.1c} 
must be imposed, as was  first observed in \cite{BHKMS}.

We now consider the supergravity-matter action 
$S= S_{\text{OMSG}} +S[V] $, where $S_{\text{OMSG}}$
denotes the action for the old minimal supergravity with a cosmological term
(see \cite{Ideas} for a review),
\bea
S_{\text{OMSG}} 
=
&=& - \frac{3}{ \k^2}
\int {\rm d}^{4} x \rd^2\q\rd^2\bar\q\,
E\,  \bar \F \F 
+ \left\{    \frac{ \m}{ \k^2} \int {\rm d}^{4} x \rd^2 \q\,
{\cal E} \,\F^3  
+ {\rm c.c.} \right\}
~,
\label{2.7}
\eea
where $\k$ is the gravitational coupling constant, and $\m$ the cosmological parameter. In the second term of \eqref{2.7}, $\cE$ denotes the chiral integration 
measure.
Varying the action $S$  with respect to the chiral compensator $\F$ 
gives the equation of motion
\bea
{\mathbb R} -{\m} = \frac{f \k^2}{6}  \F^{-2}  \big(\bar \cD^2 - 4R\big) (\bar \F V)~,
\label{2.8}
\eea
where we have introduced the super-Weyl invariant chiral scalar 
\bea
{\mathbb R} := -\frac{1}{4} \F^{-2} (\bar \cD^2 - 4R) \bar \F~.
\label{2.9}
\eea
The constraints \eqref{2.1} and the equation of motion \eqref{2.8} 
imply the nilpotency condition
\bea
({\mathbb R} -{\m})^2 =0~.
\label{2.10}
\eea
Making use of \eqref{2.8}  once more, 
the functional $S= S_{\text{OMSG}} +S[V] $
can be rewritten as the following  higher-derivative supergravity action \cite{K15}
\bea
S = \Big( \frac{3}{2f \k^2}\Big)^2  
\int {\rm d}^{4} x \rd^2\q\rd^2\bar\q\,
E \, \bar \F \F \,|{\mathbb R} -{ \m} |^2
- \left\{  \hf  \frac{ \m}{ \k^2} 
\int {\rm d}^{4} x \rd^2 \q\,
{\cal E} \,\F^3 
+ {\rm c.c.} \right\}
~,~~~
\label{2.11}
\eea
where $\mathbb R$ is subject to the constraint \eqref{2.10}.
This action is formulated purely in geometric terms, for it 
does not involve the Goldstino superfield explicitly.

The action for nilpotent old minimal supergravity, eq.  \eqref{2.11}, is universal in the sense that 
it is independent of the Goldstino superfield used to derive it.  Indeed, let us consider 
another Goldstino superfield, the most fashionable one \cite{Casalbuoni,KS}.
In curved superspace it is described by a covariantly 
chiral scalar $X$, $\bar \cD_\ad X =0$,
subject to the nilpotency condition 
\bea
X^2 =0~.
\label{2.12}
\eea
The super-Weyl invariant action for the Goldstino superfield $X$ is 
\bea
S[X , \bar X] &=& 
\int {\rm d}^{4} x \rd^2\q\rd^2\bar\q\,E\,
\bar X  X 
- \left\{ f \int {\rm d}^{4} x \rd^2 \q\,
{\cal E} \,\F^2  X
+ {\rm c.c.} \right\}
~.
\label{action12}
\eea
This action is equivalent to the one used in \cite{BFKVP,HY}.
Following \cite{K15}, we vary 
 the supergravity-matter action 
$S= S_{\text{OMSG}} + S[X , \bar X]  $ with respect to the chiral compensator $\F$, 
resulting with the equation of motion
\bea
{\mathbb R} -{\m} = -\frac{2}{3} f \k^2 \frac{X}{\F} ~.
\eea
Due to \eqref{2.12},
the equation of motion tells us that 
\eqref{2.10} holds.
Making use of \eqref{2.12}  once more, 
the action $S= S_{\text{OMSG}} + S[X , \bar X]  $ 
can be recast exactly in the form \eqref{2.11}.

The constraint \eqref{2.10} 
coincides with the one derived in \cite{BMST} within the Goldstino brane approach. 
It is also similar in form to the one postulated in \cite{DFKS}.
However,  the nilpotent supergravity action \eqref{2.11} differs from the one
used  in \cite{DFKS}. The two actions are actually related, as explained in 
Appendix D of \cite{KMcAT-M}.

The nilpotency condition \eqref{2.12} is preserved if $X$ is locally rescaled, 
$X \to \re^{\t} X$, where $\t$ is covariantly chiral,  $\bar \cD_\ad \t=0$.
Requiring the action \eqref{action12} to be stationary under such re-scalings 
gives the nonlinear equation
\bea \label{X2g}
{f} \F^2 X = -\frac 14  X ({\bar \cD}^2 -4R)\bar  X  \,.
\eea
The constraints  \eqref{2.12} and \eqref{X2g}
define the chiral Goldstino superfield $\cX$ of \cite{LR}. 

If we define $V$ to be a composite superfield, 
\bea
f  \bar \F \F V = \bar X X~,
\label{2.15}
\eea
then the  constraints \eqref{2.1} are satisfied automatically. 
Relation $f  \bar \F \F\cV = \bar \cX \cX$ holds identically 
for the irreducible Goldstino superfields
$\cV$ and $\cX$.
Plugging \eqref{2.15} into \eqref{2.4} gives
the higher derivative action
\bea
S_{\rm HD}[X, \bar X]= \int \rd^4 x \rd^2 \q  \rd^2 \bar{\q} \, E\,\left\{
\frac{1}{ 16f^2} 
\frac{ |X ( \cD^2 -4\bar R )  X |^2}
{(\bar \F \F)^2}
- 2  \bar X X \right\} ~.
\label{2.16}
\eea
Its important property is that $S_{\rm HD}[\cX, \bar \cX] = S[\cX, \bar \cX] = S[\cV]$.

Within the new minimal formulation for $\cN=1$ supergravity 
(see, e.g., \cite{Ideas} for a review),
the compensator is a  real scalar superfield ${\mathbb L}= \bar {\mathbb L}$ constrained by $(\bar \cD^2 -4R) {\mathbb L}  =0$.
We consider the supergravity-matter action 
$S= S_{\text{NMSG}} +S[V] $, where 
$S[V]$ is obtained from \eqref{2.4} by replacing $\bar \F \F \to \mathbb L$, and 
$S_{\text{NMSG}}$
is the action for new minimal supergravity
\bea
 S_{\text{NMSG}}=  \frac{3}{\k^2} \int \rd^4 x \rd^2 \q  \rd^2 \bar{\q} \, E\, 
{\Bbb L}\, {\rm ln} \frac{\Bbb L}{|\F|^2} ~,
 \label{2.18}
\eea
in which $\F$ is a purely gauge degree of freedom. 
New minimal supergravity is known to allow
no supersymmetric cosmological term. Thus
the  action $S=S_{\text{NMSG}} +S[V] $ generates a positive cosmological term.
Varying $S$
with respect to the compensator $\mathbb L$ gives the equation 
\bea
\frac{3}{2f\k^2} {\mathbb W}_\a = W_\a~, \qquad
 {\mathbb W}_\a :=  -\frac{1}{4} (\bar \cD^2 - 4R)\cD_\a {\rm ln} \frac{\Bbb L}{|\F|^2}~,
 \eea
where $W_\a$ is given by \eqref{2.2}.
This equation allows us to eliminate the Goldstino superfield 
from $S= S_{\text{NMSG}} +S[V] $, and the resulting action takes the following form
\bea
S= \Big(\frac{3}{4f\k^2} \Big)^2
\int {\rm d}^{4} x \rd^2 \q\,
{\cal E} \,{\mathbb W}^\a {\mathbb W}_\a~.
\label{2.19}
\eea
This functional is the action for $R^2$ supergravity
within the new minimal  formulation \cite{CFPS,FKP}. 
Making use of \eqref{2.3} gives
\bea
{\mathbb W}_\a =  (\bar \cD^2 - 4R)\cD_\a 
\frac{{\mathbb W}^2 \bar {\mathbb W}^2}{(\cD {\mathbb W})^3}~.
\label{2.20}
\eea
The action \eqref{2.19} and constraint \eqref{2.20}
define  nilpotent new minimal supergravity.


\section{Nilpotent $\cN=2$ reduced chiral multiplet}

The zoo of irreducible and reducible $\cN=2$ Goldstino superfields 
coupled to $\cN=2$ supergravity was described in \cite{KMcAT-M}.
A novel feature of $\cN=2 \to \cN=0$ local supersymmetry breaking
is that one can consistently define nilpotent Goldstino-matter superfields
that contain a physical gauge field (one-form or two-form) 
in addition to the two Goldstino fields 
and  some auxiliaries \cite{KT-M17}.

\subsection{Reduced chiral and linear multiplets}\label{section3.1}

In $\cN=2$ supersymmetry, 
the field strength of an Abelian vector multiplet is a {\it reduced chiral} superfield \cite{GSW}.
In curved superspace, it is a
covariantly chiral superfield $W$,  
$
\cDB^\ad_i W= 0,
$
 subject to the Bianchi identity \cite{Howe}
 \bea
\big(\cD^{ij}+4S^{ij}\big) W&=&
\big(\cDB^{ij} +4\bar{S}^{ij}\big)\bar{W} ~,
\qquad \cD^{ij}:=\cD^{\a(i}\cD_\a^{j)}~, \quad \cDB^{ij}:=\cDB_\ad^{(i}\cDB^{j) \ad}~.
\label{Bianchi}
\eea
The superfields $S^{ij} $ and ${\bar S}^{ij} $  in \eqref{Bianchi}
are special dimension-1 components of the torsion, see \cite{KLRT-M2} for the technical details of the superspace formulation
$\cN=2$  conformal supergravity  \cite{Howe} used. 
The constraints on $W$ can be solved in terms of 
Mezincescu's prepotential,
 $V_{ij}=V_{ji}$,
which is an unconstrained real $\sSU(2)$ triplet. 
The curved-superspace solution is \cite{BK11}
\begin{align}
W = \frac{1}{4}\bar\Delta \big({\cD}^{ij} + 4 S^{ij}\big) V_{ij}~.
\end{align}
Here   $\bar{\D}$ denotes the $\cN=2$ chiral projection operator 
(see, e.g.,  \cite{BK11} for the details)
\bea
\bar{\D}
&=&\frac{1}{96} \Big(\big(\cDB^{ij}+16\bar{S}^{ij}\big)\cDB_{ij}
-\big(\cDB^{\ad\bd}-16\bar{Y}^{\ad\bd}\big)\cDB_{\ad\bd} \Big)
~, \qquad \cDB^{\ad\bd}:=\cDB^{(\ad}_k\cDB^{\bd)k}~.
\label{chiral-pr}
\eea

 In curved superspace,
the $\cN=2$ tensor multiplet is  described by
its gauge-invariant field strength $G^{ij}$  which is 
a linear multiplet. The latter is 
defined to be a  real ${\sSU}(2)$ triplet (that is, 
$G^{ij}=G^{ji}$ and ${\bar G}_{ij}:=\overline{G^{ij}} = G_{ij}$)
subject to the covariant constraints  
\bea
\cD^{(i}_\a G^{jk)} =  {\bar \cD}^{(i}_\ad G^{jk)} = 0~,
\label{3.5}
\eea
which are solved in terms of a chiral
prepotential $\Psi$ (see, e.g., \cite{BK11} for the details)
\begin{align}
\label{eq_Gprepotential}
G^{ij} = \frac{1}{4}\big( \cD^{ij} +4{S}^{ij}\big) \Psi
+\frac{1}{4}\big( \cDB^{ij} +4\bar{S}^{ij}\big){\bar \Psi}~, \qquad
{\bar \cD}^i_\ad \J=0~,
\end{align}
which is invariant under Abelian gauge transformations
\bea
\d_\L \Psi = \ri \Lambda
~,
\label{3.7}
\eea
with the gauge parameter $\Lambda$ being a reduced chiral superfield.


\subsection{Deformed reduced chiral multiplet}

As defined in  \cite{K-superWeyl,KT-M15},  
a deformed 
reduced chiral superfield $\cZ$ coupled to $\cN=2$ supergravity  
 is described by the constraints
\bea\label{Z}
\bar \cD^i_\ad \cZ &=&0~, 
\qquad
\big(\cD^{ij}+4S^{ij}\big)\cZ
- \big(\cDB^{ij}+ 4\bar{S}^{ij}\big)\bar{\cZ} =4  \ri G^{ij} ~. \label{Zb}
\eea
Here $G^{ij}$  is a linear multiplet which obeys the constraints \eqref{3.5}.
In addition, $G^{ij}$ is required to be nowhere vanishing,  $G^{ij} G_{ij} \neq 0$.
We identify $G^{ij}$ with one of the two conformal compensators
of the minimal  formulation for $\cN=2$ supergravity proposed in \cite{deWPV}.

In the flat limit, a chiral superfield obeying the constraints \eqref{Zb} with 
$G^{ij} =\text{const}$ appeared in the framework
of partial $\cN=2 \to \cN=1 $ supersymmetry breaking \cite{APT,IZ1}.


\subsection{Quadratic nilpotency condition}

In \cite{KT-M15}, $\cZ$ was subject to the quadratic nilpotency condition
\bea
\cZ^2 =0~. 
\label{Zac}
\eea
The constraints  \eqref{Z} and \eqref{Zac} imply that, for certain 
$\cN=2$ supergravity backgrounds,  
the degrees of freedom described by the $\cN=2$ chiral superfield $\cZ$ 
are in one-to-one correspondence with those of an Abelian $\cN=1$ vector multiplet. 
The specific feature of such $\cN=2$ supergravity backgrounds is that 
they possess an $\cN=1$ subspace $\cM^{4|4}$ of the 
full $\cN=2$ curved superspace $\cM^{4|8}$. 
This property is not universal. In particular, there exist maximally 
$\cN=2$ supersymmetric backgrounds with no admissible truncation to $\cN=1$ 
\cite{BIL}.
As shown in  \cite{KT-M15},
the superfield constrained by  \eqref{Z} and \eqref{Zac} 
is suitable for the description of
partial $\cN=2 \to \cN=1 $ rigid supersymmetry breaking 
in every maximally supersymmetric spacetimes $\cM^{4}$
which is the bosonic body of an  $\cN=1$ superspace $\cM^{4|4} $ described 
by the following algebra of $\cN=1$ covariant derivatives\footnote{These backgrounds
are maximally supersymmetric solutions of  pure $R^2$ supergravity
\cite{Kuzenko:2016nbu}.} 
\begin{subequations}
\label{RS^3}
\bea
&\{\cD_\a,\cD_\b\}= 0~, \qquad \{\cDB_\ad,\cDB_\bd\}=0~,\qquad
\{\cD_\a,\cDB_\bd\}=-2\ri\cD_{\a\bd}~,
\\
&{[}\cD_\a,\cD_{\b\bd}{]}=\ri\ve_{\a\b}G^\g{}_{\bd}\cD_\g
~,\qquad
{[}\cDB_\ad,\cD_{\b\bd}{]}=-\ri\ve_{\ad\bd}G_\b{}^\gd\cDB_\gd~,
\\
&{[}\cD_{\a\ad},\cD_{\b\bd}{]}=
-\ri\ve_{\ad\bd}G_\b{}^\gd\cD_{\a\gd}
+\ri\ve_{\a\b}G^\g{}_\bd\cD_{\g\ad}
~,
\eea
where the real four-vector $G_{b}$ is covariantly constant,
\bea
\cD_\a G_b = 0~, \qquad G_{b} =\bar G_b~.
\eea
\end{subequations}
Since $G^2 = G^b G_b $ is constant, the geometry 
\eqref{RS^3} describes  three different superspaces, for $G_b \neq 0$, 
which correspond to the choices $G^2<0$, $G^2>0$ and $G^2=0$, 
respectively. The Lorentzian manifolds $\cM^4$ supported by these superspaces are 
${\mathbb R}\times S^3$, ${\rm AdS}_3 \times S^1$ or its covering 
${\rm AdS}_3 \times {\mathbb R}$, 
and a pp-wave spacetime
isometric to the Nappi-Witten group \cite{NappiW}, respectively.
For each of the  backgrounds \eqref{RS^3}  with $G_a \neq 0$, Ref.   \cite{KT-M15}
constructed the Maxwell-Goldstone multiplet actions
for partial $\cN=2 \to \cN=1$ supersymmetry breaking,
as a generalisation of the earlier works \cite{BG,RT} corresponding to the 
$G_a=0$ case. 
 

\subsection{Cubic nilpotency condition}

If one is interested in $\cN=2 \to \cN=0 $ breaking of local supersymmetry, 
the nilpotency condition \eqref{Zac} should be replaced with a weaker constraint 
\bea
\cZ^3 =0~.
\label{3.11}
\eea
The action for our supergravity-matter theory involves two contributions
\bea
S= S_{\rm SG}  +S[\cZ, \bar \cZ]~,
\label{3.12}
\eea
where $S_{\rm SG}$ denotes the pure supergravity action 
and $S[\cZ , \bar \cZ] $ corresponds to the Goldstino superfield.
We make use of the minimal formulation for $\cN=2$ supergravity 
with vector and tensor 
compensators \cite{deWPV}.  
In the superspace setting, the supergravity action 
can be written in the form \cite{BK11}
\bea
S_{\rm SG}  &=& \frac{1}{ \k^2} \int \rd^4 x {\rm d}^4\q \, \cE \, \Big\{
\J {\mathbb W} - \frac{1}{4} W^2 +m \J W \Big\}          +{\rm c.c.}   \non \\
 &=& \frac{1}{ \k^2} \int \rd^4 x {\rm d}^4\q \, \cE \, \Big\{
\J {\mathbb W} - \frac{1}{4} W^2 \Big\} +{\rm c.c.}
     + \frac{m}{ \k^2} \int \rd^4 x {\rm d}^4\q \rd^4\bar\theta\,
     E \, G^{ij}V_{ij}~,~~~~~~~
\label{6.11} 
\eea
where 
$m$ is the cosmological parameter.
The supergravity action involves the composite
\bea
   \mathbb W := -\frac{G}{8} (\bar \cD_{ij} + 4 \bar S_{ij}) \left(\frac{G^{ij}}{G^2} \right) ~, 
   \label{2.100}
\eea
which proves to be a reduced chiral superfield.
Eq.  \eqref{2.100} is 
one of the simplest applications of the powerful approach 
to generate composite reduced chiral multiplets presented in \cite{BK11}.

The action for the goldstino superfield $\cZ$ in \eqref{3.12} is 
\bea
S [\cZ , \bar \cZ] &=&  \int \rd^4 x {\rm d}^4\q \, \cE \, \Big\{
\frac{1}{4} \cZ^2  +\z W \cZ 
+\r \Big(\cZ \J - \frac{\ri}{2}  \J^2\Big)
\Big\}      +{\rm c.c.}   ~,
\label{3.15}
\eea
where $\z$ and $\r$ are  complex and real parameters, respectively.
The $\r$-term in \eqref{3.15} was introduced in  \cite{KT-M15},
where it was shown to be invariant under gauge transformations 
 \eqref{3.7}.

In the flat superspace limit, with $G^{ij} =\text{const}$, 
a chiral superfield $\cZ$ constrained by 
\eqref{Z} and \eqref{3.11} was considered 
in \cite{DFS}. As was demonstrated  in \cite{DFS}, 
for a certain range of parameters $G^{ij} $, 
$\cZ$ contains the following independent fields: 
two Goldstini,  a gauge one-form
and a real, nowhere vanishing, $\sSU(2) $ triplet of auxiliary fields.

Ref. \cite{KT-M17} also proposed a different $\cN=2$ Goldstino-matter multiplet
in order to describe  $\cN=2 \to \cN=0$ local supersymmetry breaking.
It is a linear superfield $\cH^{ij}$, 
$
\cD^{(i}_\a \cH^{jk)} =  {\bar \cD}^{(i}_\ad \cH^{jk)} = 0$,
which is subject to the cubic nilpotency condition 
\bea
{\cH}^{(i_1i_2} {\cH}^{i_3 i_4} {\cH}^{i_5 i_6)} =0~,
\eea
which proves to expresse the $\sSU(2)$ triplet of physical scalars, $\cH^{ij}|_{\q=0}$,
in terms of the other component fields of $\cH^{ij}$. Thus the field content of 
 $\cH^{ij}$ is as follows: two Goldstini, a gauge two-form, and a complex nowhere vanishing auxiliary scalar. The interested reader is referred to \cite{KT-M17} for the 
complete description of this model.


\noindent
{\bf Acknowledgements:} I am grateful to Ian McArthur and
Gabriele Tartaglino-Mazzucchelli for collaboration on \cite{KMcAT-M,KT-M17}
and comments on the manuscript. 
I thanks the organisers of the Workshop SQS'2017 
for their warm hospitality in Dubna.
The research presented in this  work is supported in part by the Australian 
Research Council, project No. DP160103633.


\begin{footnotesize}

\end{footnotesize}

\end{document}